\documentclass[aps,pra,doublecol,epsfigure]{epl2}
\usepackage{graphicx}
\usepackage{amsmath}    
\usepackage{latexsym}
\usepackage{amsfonts}   
\usepackage{amssymb}
\usepackage{array}      
\usepackage{epsfig}
\usepackage[colorlinks=true,linkcolor=blue,urlcolor=blue,citecolor=blue]{hyperref}
\usepackage{color}
\usepackage{hyperref,cuted}
\usepackage{braket}
\usepackage{dsfont,float} 

\DeclareMathOperator{\tr}{Tr}
\DeclareMathOperator{\diag}{diag}
\DeclareMathOperator{\Ai}{Ai}
\DeclareMathOperator{\Bi}{Bi}
\DeclareMathOperator{\arccosh}{arccosh}

\begin{document}

\title{A self-contained quantum harmonic engine}

\author{B. Reid\inst{1} \and S. Pigeon\inst{1,2} \and M. Antezza\inst{3,4} \and G. De Chiara \inst{1,3}} 
\institute{ 
\inst{1} Centre  for  Theoretical  Atomic,  Molecular  and  Optical  Physics, Queen's  University  Belfast,  Belfast  BT7 1NN,  United  Kingdom\\
\inst{2} Laboratoire Kastler Brossel, Sorbonne Universit\'e, CNRS, ENS-PSL Research University, Coll\`ege de France, 4 place Jussieu Case 74, F-75005 Paris, France\\
\inst{3} Laboratoire Charles Coulomb (L2C), UMR 5221 CNRS-UniversitŽ de Montpellier, F- 34095 Montpellier, France\\
\inst{4} Institut Universitaire de France, 1 rue Descartes, F-75231 Paris, France}
\shortauthor{B. Reid \etal}


\abstract{
We propose a system made of three quantum harmonic oscillators as a compact quantum engine for producing mechanical work. The three oscillators play respectively the role of the hot bath, the working medium and the cold bath. The working medium performs an Otto cycle during which its frequency is changed and it is sequentially coupled to each of the two other oscillators. As the two environments are finite, the lifetime of the machine is finite and after a number of cycles it stops working and needs to be reset. 
Remarkably, we show that this machine can extract more than 90\% of the available energy during 70 cycles. 
Differently from usually investigated infinite-reservoir configurations, this machine allows the protection of induced quantum correlations and we analyse the entanglement and quantum discord generated during the strokes. Interestingly, we show that high work generation is always
accompanied by large quantum correlations.
Our predictions can be useful for energy management at the nanoscale, and can be relevant for experiments with trapped ions and experiments with light in integrated optical circuits.
}

\maketitle

\section{Introduction}The recent renewed interest in quantum aspects of out-of-equilibrium thermodynamics \cite{CampisiRMP,EspositoRMP,GooldReview,AndersReview,XuerebReview,Benenti2017} has triggered an intensive research program to design and realise thermal engines whose working substance is a quantum system~\cite{AlickiJPA1979,
KosloffJCP1984,
Scully2003,
AllahverdyanPRE2005,
QuanPRE2007,
LindenPRL2010,
VenturelliPRL2013,
GelbwaserPRE2013,
SkrzypczykNatCom2014,
Correa2014,
DelCampoSR2014,
RossnagelPRL2014,
ZhangPRL2014,
MariJPB2015,
HardalSR2015,
LeggioPRA2015,
DoyeuxPRE2016,
LeggioPRE2016,
KosloffEntropy2017,
WatanabePRL2017,
NewmanPRE2017,
BissbortPRE2017,
uzdin2016zeno,
Abah2014}. Most of the proposals employ a few qubits or quantum harmonic oscillators as the working substance but a few works explore the possibility of using quantum many-body systems \cite{CampisiNatCom2016}. The interest in quantum thermal engines has received a huge thrust thanks to the experimental realisation of prototypes with trapped ions \cite{RossnagelScience,maslennikov2017quantum} and solid state devices \cite{KoskiPNAS2014,cottet2017observing}. A remarkable open problem is whether quantum effects, such as coherence, entanglement or more general quantum correlations like discord, improve the engines' performance compared to their classical counterparts. 
A unified response to the question is still lacking as many of the results found in the literature are model dependent.

\begin{figure}[t]
\begin{center}
\includegraphics[width=0.99\columnwidth]{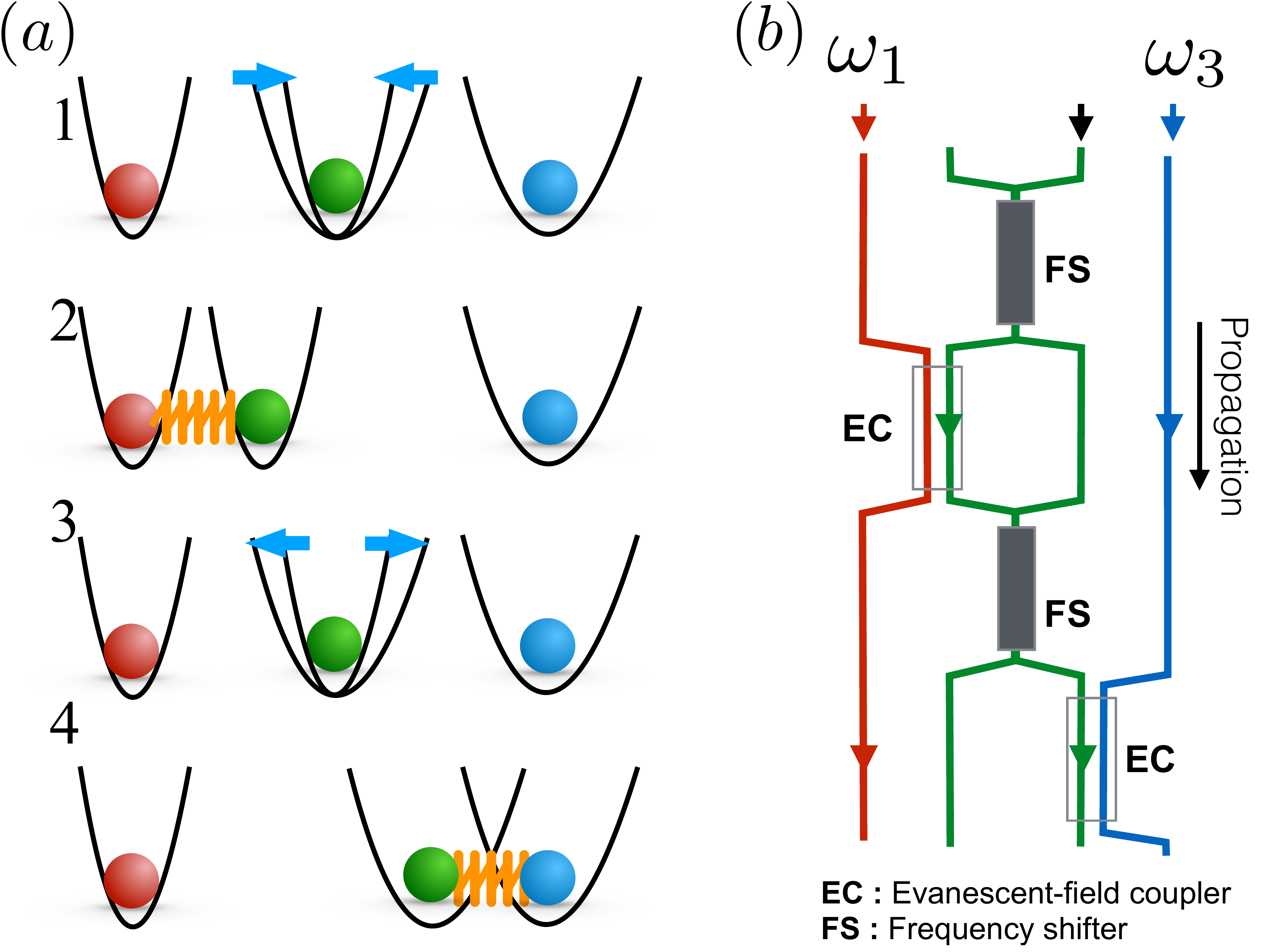}
\caption{(a) Schematic setup of the harmonic engine made of three oscillators. The central oscillator performs an Otto cycle: 1. compression, its frequency is changed from $\omega_3$ to $\omega_1>\omega_3$; 2. heating, it is coupled resonantly to the hot oscillator; 3. expansion, its frequency is changed from $\omega_1$ to $\omega_3$; 4. cooling, it is coupled resonantly to the cold oscillator. (b) Proposed experimental setup of the system utilising optical circuits. Frequency changes and oscillator interactions are performed via optical frequency shifters and evanescent coupling fields respectively.
}
\label{fig:setup}
\end{center}
\end{figure}

In all these proposals, the hot and cold environments required for the functioning of the machine are assumed to be infinite although not necessarily in equilibrium. In this work, we challenge this paradigm and consider a compact quantum engine made entirely of three quantum harmonic oscillators (see Fig.~\ref{fig:setup}(a)). One of them, the working substance, interacts sequentially with each of the other oscillators, acting as the hot and cold reservoirs, and is subject to a compression and expansion stage, thus realising the Otto cycle. Contrary to a recent proposal employing a few qubits subject to dephasing for the hot and cold reservoir~\cite{uzdin2016zeno}, the time evolution of our three-oscillator engine is always unitary.
Thermalisation of quantum systems in contact to finite structured environments has been discussed in the past
\cite{GemmerEPL2006,Benenti2015}. The modelling of open quantum system dynamics in the presence of a small environment used as a calorimeter has been studied in Ref.~\cite{SuomelaPRE2016}.

In our model, due to the finiteness and harmonicity of the environments, the working substance does not thermalise in the isochoric strokes and does not evolve in a perfect cyclic fashion. Nevertheless, we show in the following that our harmonic engine is capable of performing many cycles and producing more than 90\% of the maximum extractable work. An important consequence of the finiteness of the reservoirs is that the working substance becomes  and remains quantum correlated with the reservoirs as we show using entanglement negativity and quantum discord, which we analyse using initial thermal and squeezed states.

Our engine is self-contained and can be embedded in a larger quantum system without the need of controlled external reservoirs. Thanks to the simplicity of the model, which can be solved exactly, we believe that our engine could be implemented in any experiment with controlled quantum harmonic oscillators as, for example, trapped ions or optical modes (see Fig. \ref{fig:setup}(b)). Detailed discussion of these experimental proposals follow.

\section{Model} 
We consider three quantum harmonic oscillators of equal mass $m$ and frequency $\omega_i$ subject to the local Hamiltonians $H_{ii}=m\omega_i^2 x_i^2/2+p_i^2/2m$. The position and momentum operators for oscillator $i$ are given by $x_i=\sqrt{\hbar/2m\omega_i}(a_i+a^\dag_i)$ and $p_i=-i\sqrt{\hbar m\omega_i/2}(a_i-a^\dag_i)$, respectively, with $a_i$ ($a_i^\dag$) the annihilation (creation) operators satisfying $[a_i,a_j^\dag]=\delta_{ij}$ corresponding to the vacuum state $\ket 0_i$. We have set $\hbar=1$ everywhere.
The interaction Hamiltonian between oscillators $i$ and $j$ is 
$H_{ij}=\alpha_{ij}\left(a_i^\dag a_j+a_ia_j^\dag\right)$ characterised by the coupling strength $\alpha_{ij}$. If the oscillators are resonant ($\omega_i=\omega_j$), it follows that $[H_{ii}+H_{jj},H_{ij}]=0$ such that turning the interaction on and off does not change the energy of the system or require external work. The oscillators are initially uncorrelated and each prepared either in a thermal state with density matrix $\rho_i=e^{-\beta_i H_{ii}}/Z_i$ where $\beta_i=1/(k_BT_i)$ is the inverse temperature, $k_B$ the Boltzmann constant (we set $k_B=1$ everywhere) and $Z_i=\tr\left(e^{-\beta_i H_{ii}}\right)$ the corresponding partition function, or in a squeezed vacuum state $\ket {r_i}=\exp[r_i( a_i^2-a_i^{\dagger 2})/2]\ket 0_i$ with real squeezing parameter $r_i$.

The total Hamiltonian is a quadratic form of the oscillators' positions and momenta: $H=\sum_{ij=1}^3 H_{ij}=\sum_{\alpha,\beta=1}^6 h_{\alpha\beta} R_\alpha R_\beta$, with  $R=\{x_1,x_2,x_3,p_1,p_2,p_3\}$. As a consequence, the state of the three oscillators is always Gaussian and can be fully characterised by the covariance matrix $\sigma$, with elements: $\sigma_{\alpha\beta}=1/2\langle R_\alpha R_\beta+R_\beta R_\alpha\rangle-\langle R_\alpha\rangle\langle R_\beta\rangle$. The
 time evolution of the covariance matrix is governed by the Lyapunov equation:
\begin{equation}\label{lya}
\dot{\sigma}(t)=A\sigma(t)+\sigma(t)A^T,
\end{equation} 
  where $A=\Omega h$ and $\Omega_{\alpha\beta}=-i[R_\alpha,R_\beta]$.

\section{The cycle} 
We study the Otto cycle, described by two isentropic and two isochoric processes, or strokes. In our finite environment model oscillators 1, 2 and 3 play the role of the hot environment, the working medium and the cold environment respectively. The frequencies of oscillators 1 and 3 are constant in time $\omega_1>\omega_3$, while the frequency of the second oscillator is changed in time during the compression/expansion strokes between $\omega_1$ and $\omega_3$. Initially all couplings are zero, $\alpha_{ij}=0$.

Using the notation $E_{2i,f}^{(s)}$ to denote the energy of the second oscillator at either the initial or final time of the stroke $s$, and noticing that $E_{2f}^{(s)}\equiv E_{2i}^{(s+1)}$, the Otto cycle can be described as follows (see Fig.~\ref{fig:setup}(a)):

\begin{enumerate}
\item {\it Compression:} The frequency of the second oscillator is changed from $\omega_3$ to $\omega_1$ in a time $\tau_{\rm comp}$ according to the schedule: $\omega_2^2(t) = \omega_3^2 + (\omega_1^2-\omega_3^2) t /\tau_{\rm comp}$. The work done on the system during this stroke is defined as $W_1=E_{2i}^{(2)}-E_{2i}^{(1)}$.
 
\item {\it Heating:} The first and second oscillators are suddenly coupled for a time $\tau_H$ with coupling constant $\alpha_{12}\neq 0$, transferring energy between them. The energy transferred in this stroke is $Q_1=E_{2i}^{(2)}-E_{2i}^{(3)}$.
\item {\it Expansion:} The frequency of the second oscillator is reduced back to $\omega_3$: $\omega_2^2(t) = \omega_3^2 + (\omega_3^2-\omega_1^2) (t-\tau_ {\rm comp}-\tau_H)/\tau_{\rm exp}$ in a time $\tau_{\rm exp}=\tau_{\rm comp}$. The work done is $W_2=E_{2i}^{(4)}-E_{2i}^{(3)}$. 

\item {\it Cooling:} The second and third oscillators are suddenly coupled for a time $\tau_C$ with coupling constant $\alpha_{23}\neq 0$, transferring energy $Q_2=E^{(4)}_{2i}-E^{(4)}_{2f}$.
\end{enumerate}

The evolution of the covariance matrix during each stroke can be calculated analytically. For the heating and cooling strokes, when two oscillators are coupled, the evolution operators can be computed by performing a Bogoliubov transformation from the local coupled modes to the normal uncoupled modes. For the schedule we are considering the evolution operators for the compression and expansion strokes can be expressed in terms of Airy functions \cite{Husimi01041953}. Details are provided in the Supplementary Material. 
Two extreme cases, the instantaneous change of frequency and the very slow ramp, are particularly simple and we will discuss them in detail in the next section. 
According to our definition negative work means extracted/produced and negative heat means absorbed by the working medium.

Contrary to conventional engines, at the end of the Otto cycle it is not guaranteed that the second oscillator has returned to its initial state, so
the energy balance dictated by the first law reads:
$W_1+W_2=Q_1+Q_2+\Delta U$,
where $\Delta U=E_{2f}^{(4)}-E_{2i}^{(1)}$ is the variation of the internal energy of the second oscillator in each cycle. 
Although at each cycle the amount of work generated and heat exchanged might be different, we define a cycle efficiency as the ratio of the work  $\mathcal{W}=W_1+W_2$ produced in one cycle plus the variation of the internal energy $\Delta U$ and the heat absorbed from the two environmental oscillators:
\begin{equation}
\eta = \frac{\max(0, -\mathcal{W}+\Delta U)}{\sum_{Q_i<0}|Q_i|}
\end{equation}
Thus, the machine has a positive efficiency if it produces work or if it accumulates energy in the working medium (like a battery) to be used in a subsequent cycle. Thanks to the energy balance $0\le \eta\le 1$ and $\eta=1$ when both $Q_1$ and $Q_2$ are negative, i.e. no heat is released. In the standard case with infinite reservoirs $Q_1<0$ and $Q_2>0$ we obtain the more common formula: $\eta = 1-Q_2/|Q_1|$.

Looking at the whole system of three quantum oscillators as a driven system initially in a non passive state (the product of local thermal passive states is not necessarily passive), we can assess the performance of the process by looking at the ratio of the work done, if any, and the ergotropy $\varepsilon$, which we compute numerically \cite{erg,FuscoPRE2016}. Erogotropy is described as the maximum amount of work that can be extracted from a quantum system through a unitary operation. Briefly, consider the initial state of a system $\rho_i=\sum_jq_j\ket{q_j}\bra{q_j}$ with Hamiltonian $H=\sum_j\epsilon_j\ket{\epsilon_j}\bra{\epsilon_j}$ where we assume the ordering $q_j\geq q_{j+1}$ and $\epsilon_j\leq \epsilon_{j+1}$. With this ordering, the least energetic state that can be obtained with a unitary operation can be written as $\rho_f=\sum_jq_j\ket{\epsilon_j}\bra{\epsilon_j}$. The ergotropy can then be defined as the work extracted in the process:\begin{equation}
\varepsilon=\sum_{ij}q_j\epsilon_i\left(\left|\braket{q_j|\epsilon_i}\right|^2-\delta_{ij}\right).
\end{equation}

\section{One cycle} 
\noindent We start our analysis with the case in which the interactions between the oscillators are very small thus mimicking the weak coupling regime of open quantum systems. In this regime we can expand all our expressions up to second order in $\alpha_{ij}\tau_{H,C}$.
We start with the case in which the compression/expansion strokes are instantaneous. We also assume that the second and third oscillator are initially thermalised ($\beta_2=\beta_3$) for thermal states or have the same initial squeezing parameter ($r_2=r_3$) for squeezed states. Under these assumptions the total work obtained after one cycle for thermal states is given by:
\begin{equation}\label{workanathm}
\mathcal{W}_{th}=\frac{\tau_{H}^2\left(\omega_1^2-\omega_3^2\right)}{4\omega_1\omega_3}\left[\omega_1(\alpha_{12}^2+\omega_1^2-\omega_3^2)c_3-\alpha_{12}^2\omega_3 c_1\right],
\end{equation}
where for ease of notation we have written $c_i=\coth\left(\beta_i\omega_i/2\right)$. The equivalent expression for initially squeezed states is given by:
\begin{align}\label{workanasqz}
\nonumber&\hspace{-30pt}\mathcal{W}_{sq}=\frac{\tau_{H}^2\left(\omega_1^2-\omega_3^2\right)}{4\omega_1\omega_3}\times\\
&\left[\omega_1(\alpha_{12}^2+\omega_1^2-e^{4r_3}\omega_3^2)e^{-2r_3}
-\alpha_{12}^2\omega_3 e^{2r_1}\right].
\end{align}
In order for work to be extracted, we require Eqs.~(\ref{workanathm}, \ref{workanasqz}) to be negative. In the limit of zero initial temperature (similarly zero squeezing) in oscillators two and three, the special case we will consider here, the following inequality must be satisfied:
\begin{equation}\label{negwork}
X>\frac{\omega_1}{\omega_3}\left(1+\frac{\omega_1^2-\omega_3^2}{\alpha_{12}^2}\right),
\end{equation}
where $X=\coth\left(\beta_1\omega_1/2\right)$ for initially thermal states and $X=e^{2r_1}$ for initially squeezed states. This inequality holds for sufficiently large temperatures $T_1$ or for large initial squeezing $r_1$. Since $X$ represents the excitation energy of the first oscillator, inequality \eqref{negwork} is a requirement on the initial available energy of the hot oscillator. 

For these two cases we address the entanglement generated during the heating stroke between the first and second oscillator. This is the largest entanglement created during the process since the following strokes deteriorate or destroy it. For Gaussian states it is convenient to use the logarithmic negativity as a measure of entanglement \cite{VidalWernerPRA2002}. For two oscillators $i$ and $j$, this is defined as $E_{ij}=\max(0,-\log|2\nu_{ij}|)$ where $\nu_{ij}$ is the smallest symplectic eigenvalue of the partially transposed covariance matrix: $\sigma^{T_j}=P\sigma_{ij}P$, where $\sigma_{ij}$ is the restriction of the full covariance matrix to the oscillators $i$ and $j$ and $P=\diag(1,1,1,-1)$. Notice that entanglement is nonzero only if $\nu_{ij}<1/2$.
\begin{figure}[t]
\begin{center}
\includegraphics[width=0.49\columnwidth]{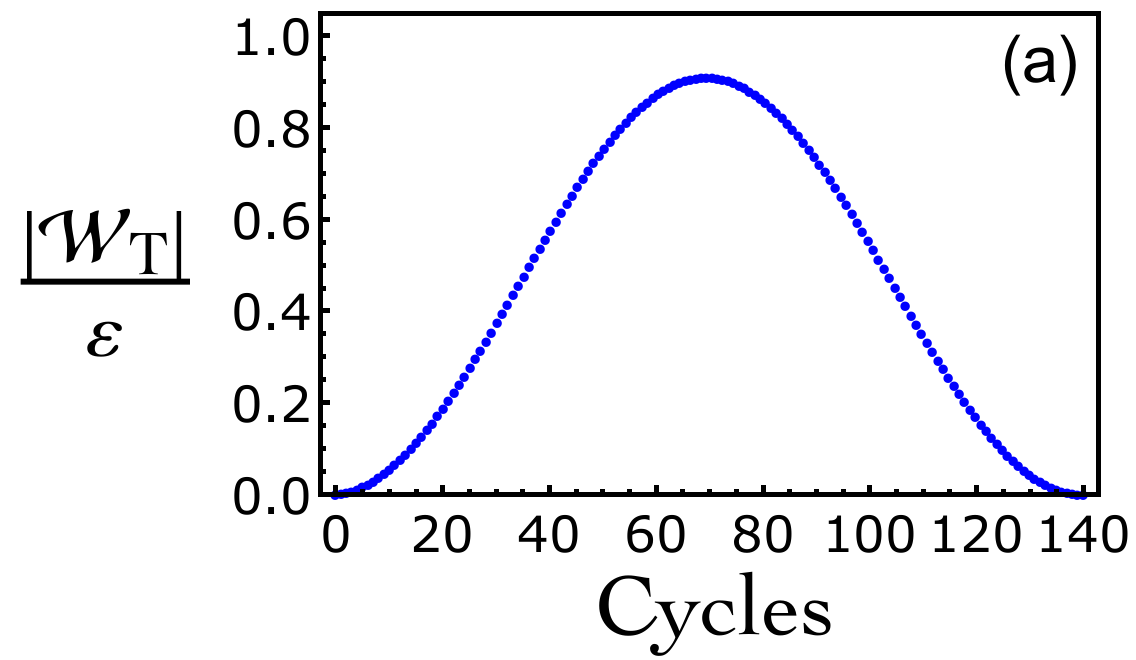}
\includegraphics[width=0.49\columnwidth]{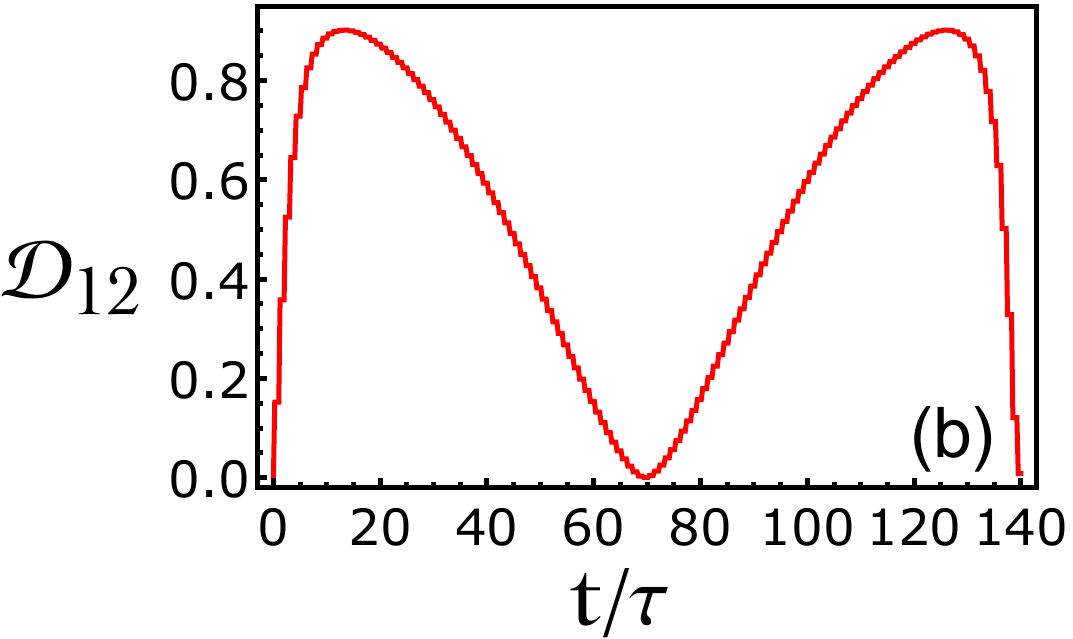}
\caption{(a) The ratio between the total work accumulated $\mathcal{W}_T$ and the ergotropy $\varepsilon$ of the initial state for each cycle in the optimised case. The parameters are: $\omega_3=0.1\omega_1, \tau_{\rm comp}=85.02\omega_1^{-1}, \tau_H=0.59\omega_1^{-1}, \tau_C=0.9996\omega_1^{-1},\alpha_{12}=0.038\omega_1,\alpha_{23}=10^{-4}\omega_1$. (b) The Gaussian discord dynamics between the first and second oscillator, $\mathcal{D}_{12}$, with these parameters.
}
\label{fig:optimal}
\end{center}
\end{figure}

When the oscillators are initially prepared in thermal states we obtain:
\begin{align}
\vspace{-10pt}
\nonumber&\nu_{12}=\\
&c_3\left[\frac{(1+A)\omega_1\omega_3c_1^2-A(\omega_1^2+\omega_3^2)c_1c_3-(1-A)\omega_1\omega_3c_3^2}{2\omega_1\omega_3(c_1^2-c_3^2)}\right]
\end{align}
with $A=2\left(\alpha_{12}\tau_H\right)^2$.
If we consider $T_3=0$ such that $c_3\rightarrow1$, the condition for entanglement generation ($\nu_{12}<1/2$) becomes $\coth\left(\beta_1\omega_1\right)<(\omega_1^2+\omega_3^2)/(2\omega_1\omega_3)$ which happens for low enough temperatures. However, due to Eq.~\eqref{negwork}, there is no work produced. Thus for initial thermal states, work extraction and entanglement are incompatible. As we will see later, however, other forms of quantum correlations might arise in the process.

If the first oscillator is instead in an initial squeezed state the symplectic eigenvalue $\nu_{12}$ takes the form:
\begin{equation}
\nu_{12}=\frac{1}{2}-\frac{\sqrt{A}e^{-(r_1+r_3)}\left|\Phi\right|}{2\sqrt{2\omega_1\omega_3}}+\frac{Ae^{-2(r_1+r_3)}\left|\Phi\right|^2}{8\omega_1\omega_3},
\end{equation} where $\Phi=\left(\omega_1-e^{2(r_1+r_3)}\omega_3\right)$. 

For quasi-static compression/expansion, using the asymptotic formulae presented in the Supplementary Material, we find that work is always extracted after the first cycle:
\begin{equation}
\mathcal{W}_{th}=-\frac{1}{2}\alpha_{12}^2\tau_H^2\left(\omega_1-\omega_3\right)\left(c_1-c_3\right),
\end{equation}
if we have $c_1>c_3$, when the oscillators are initially prepared in thermal states. For oscillators prepared in initially squeezed states, the expression for work extraction after the first cycle is given by:
\begin{equation}
\mathcal{W}_{sq}=-\frac{\alpha_{12}^2\tau_H^2(\omega_1-\omega_3)}{4}\left(e^{2r_1}-e^{2r_3}\right)\left(1-e^{-2(r_1+r_3)}\right),
\end{equation}which is negative for $r_1>r_3$.

The symplectic eigenvalue between the first and second oscillator for the quasi-static case is given by, for thermal states, 
\begin{equation}
\nu_{12}=\coth\left(\frac{\beta_3\omega_3}{2}\right)\left\{\frac{1}{2}-\alpha_{12}^2\tau_H^2\left[\frac{\sinh\left(\frac{\beta_1\omega_1-\beta_3\omega_3}{2}\right)}{\sinh\left(\frac{\beta_1\omega_1+\beta_3\omega_3}{2}\right)}\right]\right\}.
\end{equation} Taking $T_3=0$, for which we would expect a higher value of entanglement between the first and second oscillators, the limit of this function is strictly greater than $1/2$, so no entanglement is generated.

For initially squeezed states the expression for the lowest symplectic eigenvalue is too lengthy to report here. However, preparing the second and third oscillator with zero initial squeezing, entanglement generation is guaranteed between the first and second oscillator: $\nu_{12} = \frac 12 -\alpha_{12}\tau_H \sinh{r_1}$
for non-zero $r_1$. To summarise, to get entanglement created in the initial heating stroke it is sufficient, but not necessary, to prepare the first oscillator in a squeezed state.

In all the cases discussed under the weak coupling approximation, we find that $Q_2=0$ and thus, if $\mathcal{W}<0$ , $\eta=1$. However it is important to stress that since these analytical calculations refer to a single cycle, the work extracted to ergotropy ratio is very small. In the following we discuss our numerical results in a more general scenario.

\section{Many cycles} 
To fully characterise the performance of the engine over many cycles we use numerical optimisation to enhance work extraction in our machine.
We set $\beta_1=10^{-2}\omega_1^{-1}$, $\beta_{2,3}\rightarrow \infty$ and $\omega_1=1$.  For values of $0<\omega_3<\omega_1$, the variables to be optimised are the coupling values $\alpha_{ij}$ and corresponding coupling times $\tau_{H,C}$ as well as the compression/expansion time $\tau_{\rm comp}$. To this end, we employ the built-in Mathematica function NMinimize \cite{wolfram}. To avoid large interaction times and large coupling values, which would result in a short-lived engine, each parameter was restricted to a finite interval: $10^{-4}\leq\alpha_{ij}\omega_1^{-1}\leq0.05$, $10^{-3}\leq\omega_1\tau_{H,C}\leq1$, $1\leq\omega_1\tau_R\leq100$. 

For $\omega_3=0.1\omega_1$, the result of the optimisation is shown in Fig.~\ref{fig:optimal}(a) where the total work extracted to ergotropy ratio is shown. The machine extracts work until the 70th cycle and then reverses itself absorbing work and dumping heat into the first oscillator, returning almost completely to its initial state. At the end of the 70th cycle the total work extracted from the engine is $\mathcal W_T\simeq -89.5\omega_1$,  approximately $91\%$ of the ergotropy of the initial state. It should be noted that the parameters reported in the caption of Fig.~\ref{fig:optimal} are not unique and other sets of parameters can provide a similar amount of work extraction.

Due to the small value of $\alpha_{23}$ we find that the heat exchanged to the third oscillator is negligible: $Q_2\sim 0$, thus the thermal efficiency of the machine is practically 1. For an optimised engine such as the one presented here, $Q_2$ will always be small as most of the energy will be extracted during the expansion stroke of the cycle. This feature is specific to an optimised engine: there exist other sets of parameters, leading to a functioning engine, albeit with a smaller value of extracted work, that have larger values of $\alpha_{23}$ giving $Q_2$ small but non-zero. It is interesting that we obtain the same result for the total work extracted, within our working precision, by eliminating the third oscillator altogether. This is a consequence of the non-cyclic nature of the engine.

The choice of preparing the first and second oscillators in their vacuum states is not accidental, we found this case allows for the highest work to ergotropy ratio. However, non-zero initial values of $T_{2,3}$ also allows for the extraction of work. In fact, we found that for $T_2=T_3>0$, and disallowing instances where the second oscillator can ``ignore'' the third with small coupling values and/or times (such as the case presented above) optimal work extraction decreases linearly with $T_3$. Setting $T_3=0$ and $T_2>0$ we found that maximum work extraction is identical to that presented above: $\mathcal{W}_T\simeq-89.5\omega_1$. However, the work to ergotropy ratio is lower.

 As shown in the previous section entanglement between initially thermal states is not present in a regime where work can be extracted, and so we evaluate the non-classicality of this process via the Gaussian quantum discord \cite{ZurekDiscord2001,HendersonVedral2001, IlluminatiDiscord2004, AdessoDiscord2010, Olivares2012,Steve2016}. We provide a brief, mathematical definition of Gaussian quantum discord in the Supplementary Material. Figure ~\ref{fig:optimal}(b) shows the discord dynamics between the first and second oscillator as a function of time. It is symmetric about the point of reversal for the engine and is the only non-classical correlation present in the engine with $\mathcal{D}_{23}$, $\mathcal{D}_{13}=0$.  Detailed information on the internal energies of the three oscillators performing many cycles can be found in the Supplementary Material.


Figure~\ref{fig:optimal2} shows optimised work extraction for changing values of $\omega_3$. The maximum amount of extractable work, as a proportion of the ergotropy of the initial state, decreases linearly with increasing $\omega_3$, until it disappears for $\omega_3\sim\omega_1$. While work extraction is possible for any value of $\omega_3<\omega_1$, clearly our machine operates better in the regime $\omega_3\ll\omega_1$.

\begin{figure}[t]
\begin{center}
\includegraphics[width=0.7\columnwidth]{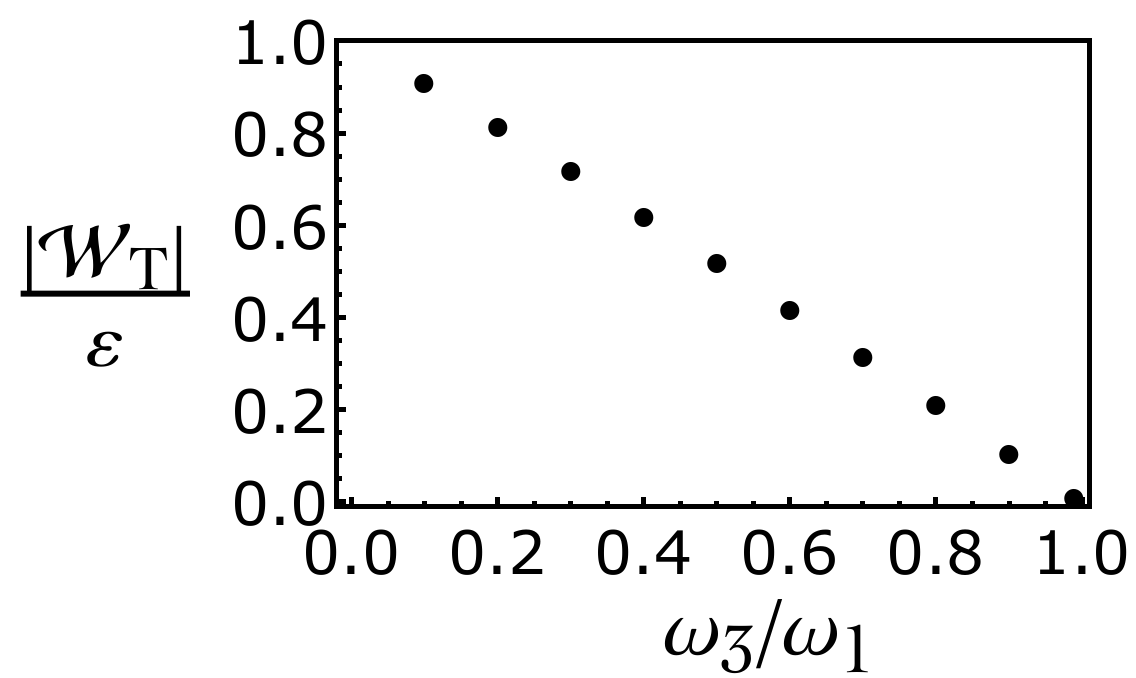}
\caption{
Ratio between optimised total extracted work and ergotropy as a function of $\omega_3$. 
}
\label{fig:optimal2}
\end{center}
\end{figure}

\section{Quantum correlations} 
The optimised case presented above provides an insight on the performance of the engine for maximum work production. While this is arguably the most important aspect of any machine, it is unclear on what connection, if any, exists between work extraction and non-classical correlations in the system. In order to investigate any link between thermodynamic quantities (work, efficiency) and non-classical correlations (entanglement, discord) we simulate the engine working with random parameters. 

We consider two cases: initially thermal states and initially squeezed states. We fix the frequency of the first oscillator ($\omega_1=1$) and the initial temperatures (squeezing parameters) for thermal (squeezed) states. The variables to be randomized are $\omega_3$, the coupling strengths $\alpha_{ij}$ and times $\tau_{H,C}$ and $\tau_{\rm comp}$. The engine is prepared with these random variables and performs Otto cycles until the criteria $\mathcal{W}<0$ is no longer satisfied. This process is repeated ten thousand times. For thermal states, we have chosen variables $\beta_1=10^{-2}\omega_1^{-1}$ and $\beta_{2,3}\rightarrow\infty$. For squeezed states, we fix $r_{2,3}=0$ and $r_1=1/2\arccosh\big[\coth(\beta_1\omega_1/2)\big]$ so the initial energy of the first oscillator is equal in both cases.

For initial thermal states, there is no significant generation of entanglement and we assess the amount of quantum correlations between any two oscillators using quantum discord. For initial squeezed states, instead we show entanglement measured by the logarithmic negativity defined above.

For thermal states, Fig.~\ref{fig:quantumcorrelations}(a-b-c) show the scatter plots of the maximum discord generated between any two oscillators at the end of a cycle and the total work produced with a given set of parameters. The discord $\mathcal D_{12}$ between the first two oscillators show a clear pattern: high values of work are necessarily accompanied by large quantum discord. In order to produce work, there must be a coherent energy transfer between the first and second oscillator that can only happen if the two oscillators become correlated. This is also reminiscent of synchronisation phenomena in coupled harmonic oscillators~(see for example~\cite{Shim2007,GiorgiPRA2012, ManzanoSR2013,MariPRL2013,WalterPRL2014}). In contrast, the quantum discords of the other pairs of oscillators do not show such a connection, and large values of work are not necessarily followed by large values of discord. Notice also that the discord $\mathcal D_{13}$ created between the first and third oscillator is mediated through the central oscillator. 

For initial squeezed states, the results for the maximum negativity created are shown in Fig.~\ref{fig:quantumcorrelations}(d-e-f). As in the case of initial thermal states, a strong entanglement between the first two oscillators is necessary for extracting work. There also seems to be a maximum value of entanglement that can be created  connected to the initial amount of squeezing of the first oscillator. 
Entanglement between the second and third oscillator and between the first and third oscillator tends to be low and correlated to small values of total work. This suggests a minimal involvement of the third oscillator in the work extraction.

\section{Implementation} Our self-contained engine can be implemented in any experiment with controlled quantum harmonic oscillators. Two platforms fit especially well our proposal: trapped ions and integrated optical circuits. With trapped ions, the role of each oscillator is played by the ion displacement from equilibrium. The trapping frequency and the ions distance, determining their interactions, can be accurately controlled \cite{Brown2011}. Thanks to the advance in state engineering,  it is possible to prepare the ions in thermal and squeezed states \cite{Meekhof1996}.

Integrated optical circuits are another interesting candidate. The time evolution of a quantum oscillator can be mapped to the propagation of an electromagnetic field in a single mode nanofibre. Preparation of thermal and squeezed states has been well mastered in quantum optics \cite{Arecchi1965,Walls1983}. The circuit proposal in Fig. \ref{fig:setup}(b) allows optical states to undergo frequency shifting and coupling via evanescent fields which realise $H_{ij}$ \cite{Crespi2013}. The main difficulty with this system is to shift the frequency of the oscillators. For frequency shifting based on acoustic-optic effect only a modulation of a few percent can be reached with high efficiency \cite{Birks1996}. Frequency conversion is an active field of research and new methods based on doped fibres have been recently developed, allowing for a larger frequency modulation but still with smaller efficiency \cite{Xu1993,Chou1999,Bouchier2005}.

\section{Conclusions}  
\begin{figure*}[t]
    \centering
   \includegraphics[width=0.52\columnwidth]{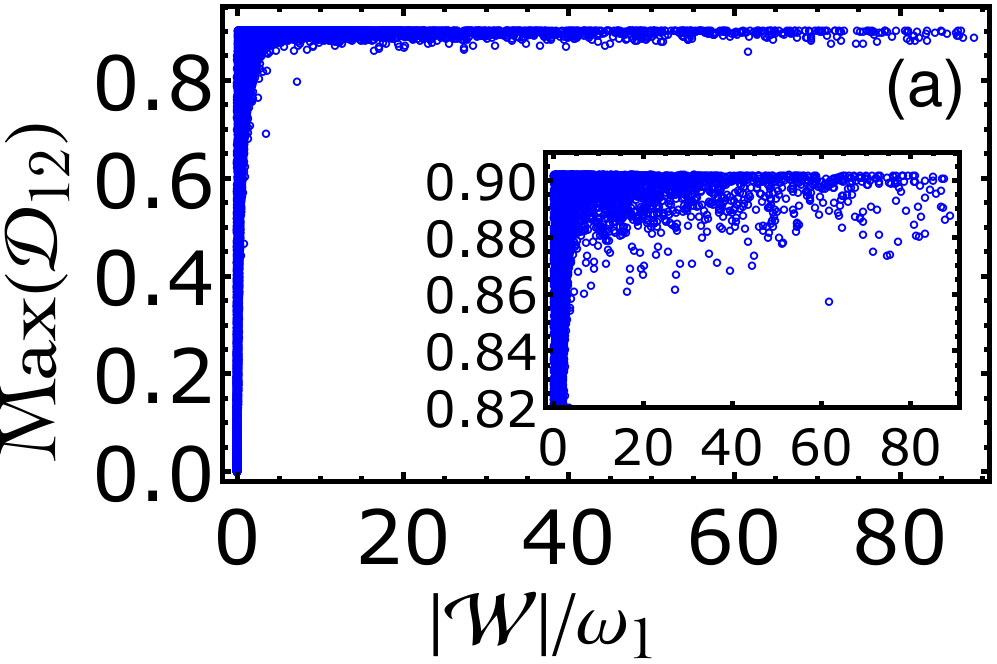}
   \includegraphics[width=0.52\columnwidth]{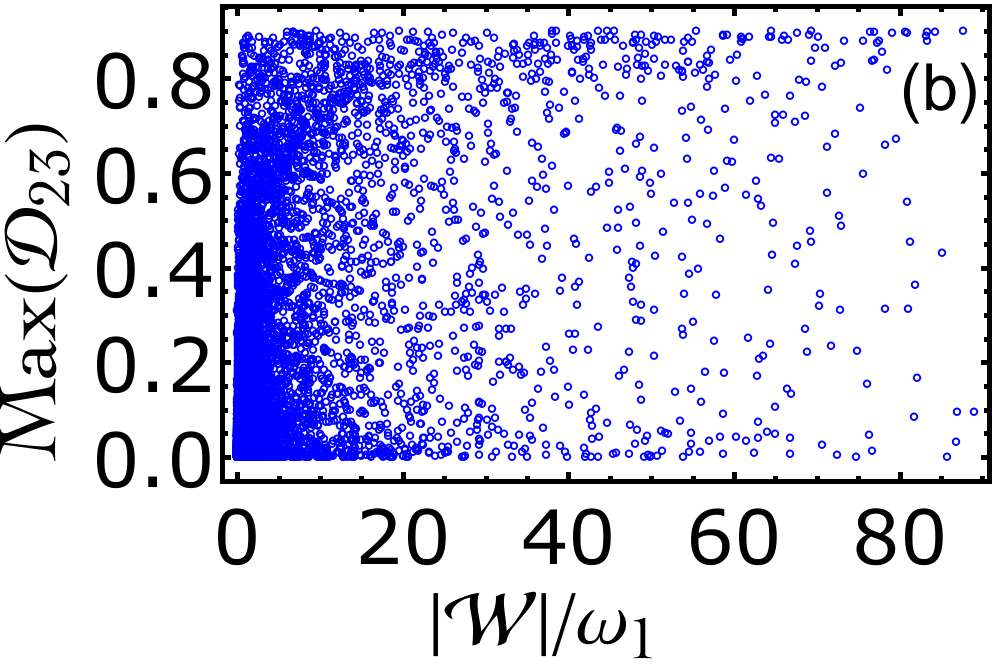}
   \includegraphics[width=0.52\columnwidth]{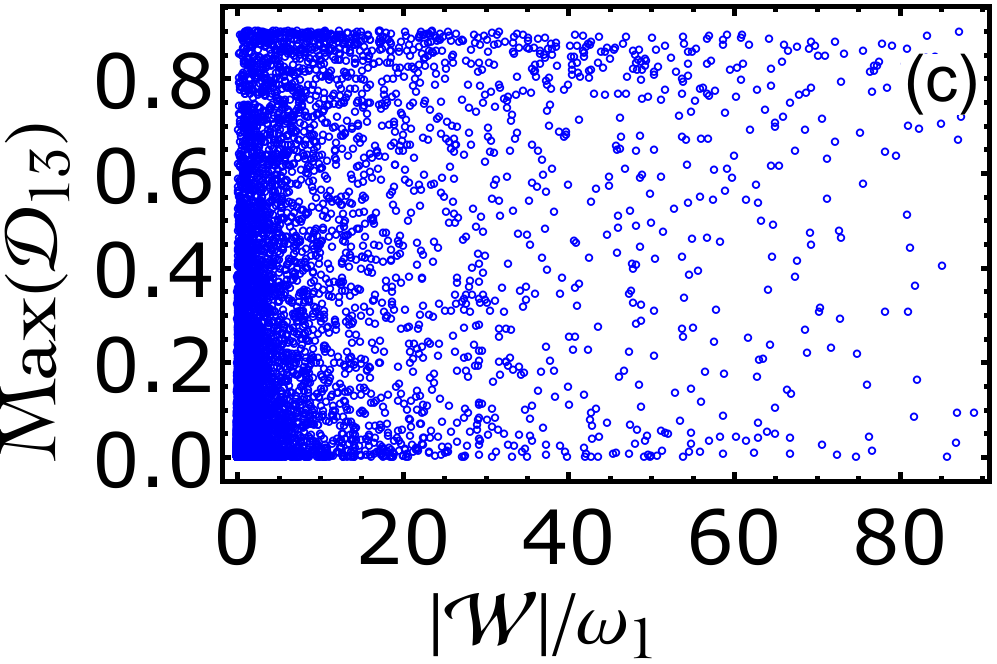}
   \\
    \includegraphics[width=0.52\columnwidth]{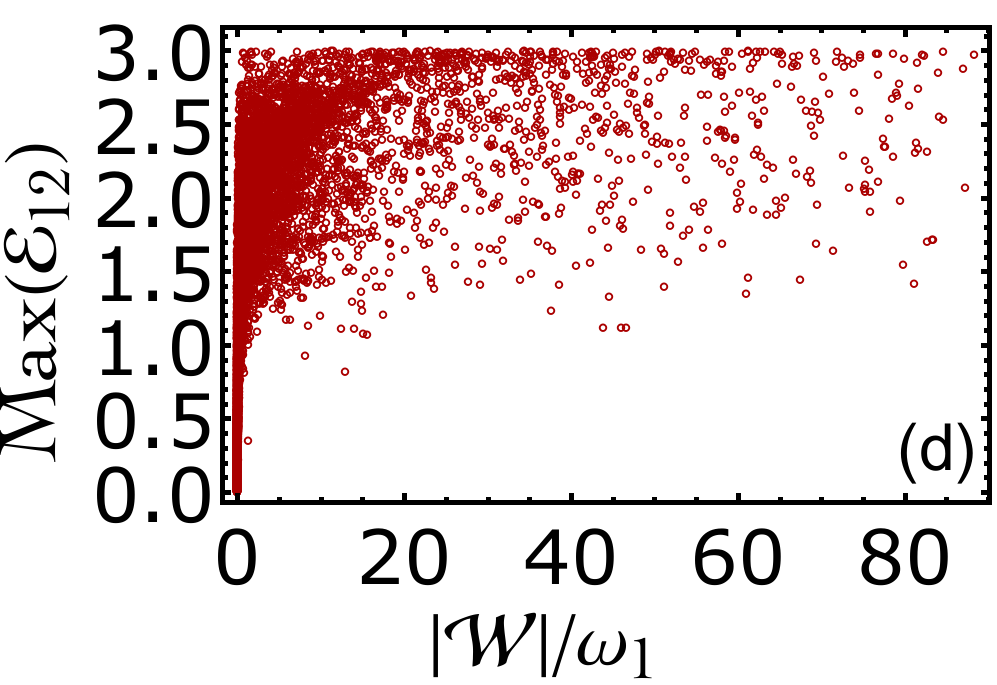}
   \includegraphics[width=0.52\columnwidth]{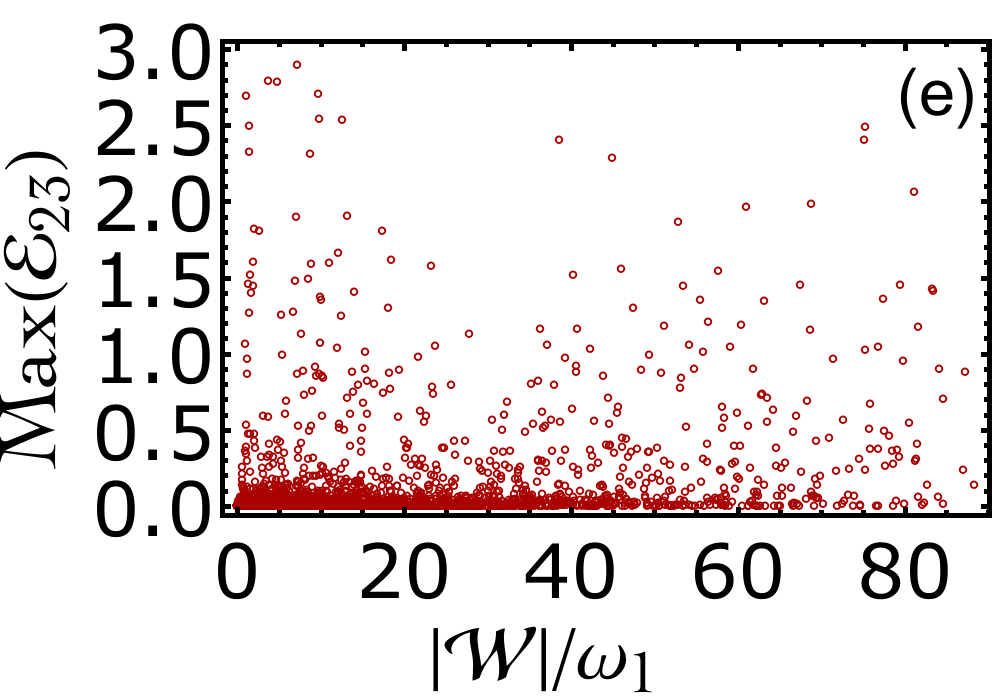}
   \includegraphics[width=0.52\columnwidth]{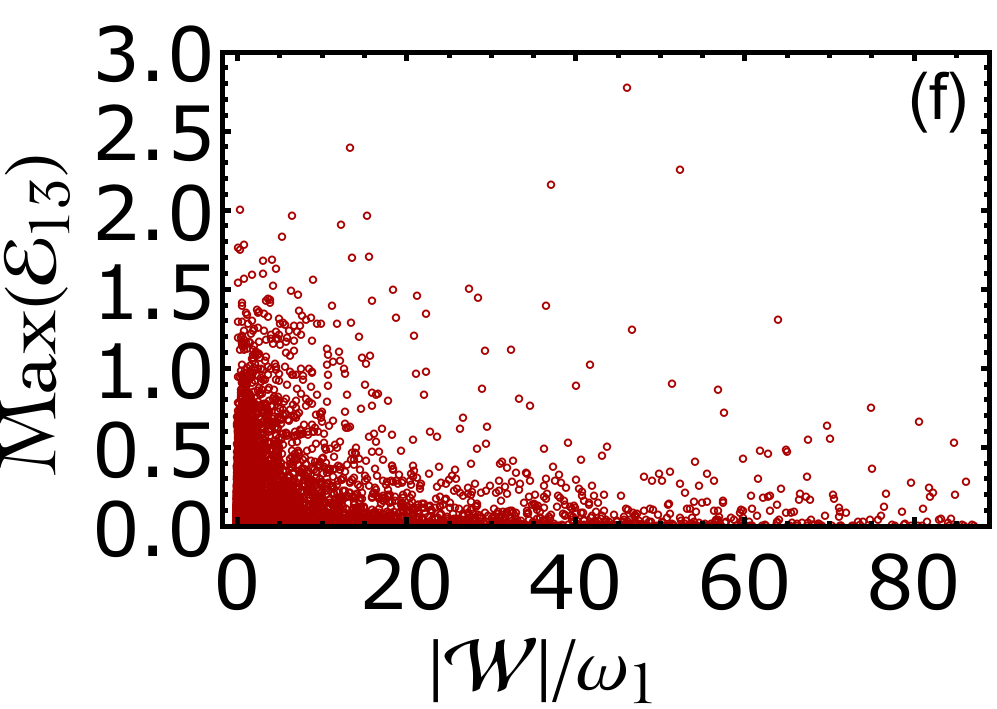}
    \caption{Scatter plots of quantum correlations versus the total work extracted in thermal machines with random parameters. Panels (a-b-c) show the maximum quantum discord between the first and second oscillators (a), second and third (b) and first and third (c).
    Panels (d-e-f) show the maximum logarithmic negativity between the first and second oscillators (d), second and third (e) and first and third (f).
 }
    \label{fig:quantumcorrelations}
\end{figure*}

In this work we have designed the simplest quantum engine made of three quantum harmonic oscillators and performing the Otto cycle. This machine does not require the use of external infinite thermal reservoirs but it operates thanks to an initial non-passive state. As such it can be integrated in a larger quantum machine that needs mechanical work in a very localised region of space. Since its resources are finite so is its lifetime. However we have shown that the optimised engine works for about 70 cycles before stopping, extracting approximately 91\% of the available energy. Our investigation confirms the common belief, although not conclusively proven, that quantum correlations enhance work extraction as evidenced numerically in our Fig.~\ref{fig:quantumcorrelations}.

Our research opens a new avenue in the study of quantum thermodynamic devices with finite reservoirs. Extending the lifetime of the engine could be achieved by coupling more oscillators onto the first and third oscillators, thus increasing the size of the energy reservoirs \cite{Karen2017}. It has also been shown that systems coupled to a finite reservoir can mimic the action of a system connected to an infinite reservoir \cite{Esposito2017}. Finally, it would be interesting to extend a program such as the one we presented here to spin systems and many-body lattice systems.
\acknowledgments
We acknowledge illuminating discussions with O. Abah, A.~Ferraro, A. Hewgill, V. Parigi, J. P. Paz and A. J. Roncaglia. GDC thanks the CNRS and the group ÒTheory
of Light-Matter and Quantum PhenomenaÓ of the Laboratoire
Charles Coulomb for hospitality during his stay in Montpellier. BR would like to thank S. Campbell for illuminating discussions.
This work was supported by the EU Collaborative project TherMiQ (grant agreement 618074), the John Templeton Foundation (grant ID 43467) the French ANR (ACHN C- Flight), the UK EPSRC and the Professor Caldwell Travel Scholarship.

\bibliographystyle{eplbib}
\bibliography{biblio}
\onecolumn

\begin{centering}
\hspace{150pt}{\LARGE {\bf \underline{Supplementary Material}}}
\end{centering}
\vspace{25pt}
\section{Evolution operators for the various strokes}
\noindent In this section we find analytically the evolution operators needed to evolve the covariance matrix in time. To this end we need to solve the Lyapunov equation $\dot{\sigma}(t)=A\sigma(t)+\sigma(t)A^T$. For a time independent Hamiltonian, i.e. the matrix $A$ is constant, the solution is simply:
\begin{equation}
\sigma(t) = \exp(At) \sigma(0) \exp(A^T t)
\end{equation}
This is the case of the heating and cooling strokes when two resonant oscillators are coupled. 
When the first and second oscillators, both at frequency $\omega_1$, are coupled $\alpha_{12}\neq 0$ and $\alpha_{23}=0$. The matrix $A$ reads:
\begin{equation}
A=\left(\begin{array}{cccccc}
0 & 0 & 0 & 1 & \alpha_{12}/\omega_1 & 0 \\
0 & 0 & 0 & \alpha_{12}/\omega_1 & 1 & 0\\
0 & 0 & 0 & 0 & 0 & 1 \\
-\omega_1^2 & -\alpha_{12} \omega_1 & 0 & 0 & 0 & 0 \\
-\alpha_{12}\omega_1 & -\omega_1^2 & 0 & 0 & 0 & 0 \\
0 & 0 & -\omega_3^2 & 0 & 0 & 0
\end{array}\right)
\end{equation}
and we obtain:
\begin{align}
\nonumber\hspace{-40pt} &\exp(At)= \\
 &\left(\begin{array}{cccccc}
\cos\alpha_{12}t\cos\omega_1t & -\sin\alpha_{12}t\sin\omega_1t & 0 &1/\omega_1 \cos\alpha_{12}t\sin\omega_1 t & 1/\omega_1 \sin\alpha_{12}t\cos\omega_1 t & 0 \\
-\sin\alpha_{12}t\sin\omega_1t    &  \cos\alpha_{12}t\cos\omega_1t& 0 & 1/\omega_1 \sin\alpha_{12}t\cos\omega_1 t & 
1/\omega_1 \cos\alpha_{12}t\sin\omega_1 t & 0 \\
0 & 0 & \cos\omega_3t & 0 & 0 & 1/\omega_3 \sin\omega_3 t \\
-\omega_1\cos\alpha_{12}t\sin\omega_1t & -\omega_1\sin\alpha_{12}t\cos\omega_1t & 0 & \cos\alpha_{12}t\cos\omega_1t & -\sin\alpha_{12}t\sin\omega_1 t & 0 \\
-\omega_1\sin\alpha_{12}t\cos\omega_1t & -\omega_1\cos\alpha_{12}t\sin\omega_1t & 0 &-\sin\alpha_{12}t\sin\omega_1 t & \cos\alpha_{12}t\cos\omega_1t &  0 \\
0 & 0 & -\omega_3\sin\omega_3t & 0 & 0 & \cos\omega_3t
\end{array}\right)
\end{align}
A similar expression for $A$ and $\exp A t$ is obtained as well when $\alpha_{23}\neq 0$, $\alpha_{12}=0$ and $\omega_2=\omega_3$. 

Now let us consider the compression/expansion strokes. In this case the oscillators are uncoupled and the second oscillator is subject to a time-dependent frequency. The reordered Hamiltonian matrix is:
\begin{equation}
A(t) =\left(\begin{array}{cc}
0_3 & \mathds{1}_3 \\
-\omega^2(t) &0_3
\end{array}\right)
\end{equation}
where $0_3$ and $\mathds{1}_3$ are the $3\times 3$ zero and identity matrices, respectively, while $\omega^2(t)=\diag(\omega_1^2,\omega_2^2(t),\omega_3^2)$.
The solution to the Lyapunov equation (1) is thus $\sigma(t)=U\sigma(0)U^T$, where:
\begin{equation}
U = \left(\begin{array}{cc}
a & b \\
c &d
\end{array}\right)
\end{equation}
and 
\begin{eqnarray}
a&=& \diag(\cos\omega_1t, y(t), \cos\omega_3 t)
\\
b&=& \diag(1/\omega_1\sin\omega_1t, x(t), 1/\omega_3\sin\omega_3 t)
\\
c&=&\dot a    \\    d&=&\dot b
\end{eqnarray}
The functions $x(t)$ and $y(t)$ fulfil the equations:
\begin{eqnarray}
\ddot x(t)&=&-\omega_2^2 x(t), \quad x(0)=0,\quad \dot x(0)=1
\\
\ddot y(t)&=&-\omega_2^2 y(t), \quad y(0)=1,\quad \dot y(0)=0
\end{eqnarray}
These equations can be solved analytically for the schedule 
$\omega_2^2(t) = \omega_{\rm in}^2 + (\omega_{\rm fin}^2-\omega_{\rm in}^2) t /\tau$. Their solutions are given in terms of Airy functions:
\begin{eqnarray}
x(t)&=& -\pi\left(\frac{\tau}{\omega_{\rm in}^2-\omega_{\rm fin}^2}\right )^{1/3}\left\{\Ai[z(t)]\Bi(w)-\Ai(w)\Bi[z(t)]\right\}
\\
y(t)&=& -\pi\left\{\Bi[z(t)]\Ai'(w)-\Ai[z(t)]\Bi'(w)\right\}
\end{eqnarray}
where 
\begin{eqnarray}
w&=&-\omega_{\rm in}^2 \left[\frac{\tau}{\omega_{\rm in}^2-\omega_{\rm fin}^2}\right]^{2/3}
\\
z(t)&=&-\omega_2^2(t)\left[\frac{\tau}{\omega_{\rm in}^2-\omega_{\rm fin}^2}\right]^{2/3}
\end{eqnarray}

In our calculations we specialise to two extreme cases, one in the limit $\tau\to 0$ in which case we simply have $U=\mathds{1}_6$ and the other for very slow ramp: $\tau\gg \max(\omega_{\rm in}^{-1},\omega_{\rm fin}^{-1})$. In this case we can use the asymptotic expansion of the Airy functions for large arguments. Under this approximation the evolution operator takes the form:
\begin{equation}
\label{eq:Uasym}
U=\left(\begin{array}{cccccc}
\cos\omega_1\tau & 0 & 0 & 1/\omega_1\sin\omega_1\tau &0 & 0 \\
0 & \sqrt{\frac{\omega_{\rm in}}{\omega_{\rm fin}}}\cos\phi & 0 & 0 & \frac{\sin\phi}{\sqrt{\omega_{\rm in}\omega_{\rm fin}}} & 0\\
0 & 0 & \cos\omega_3\tau & 0 & 0 & 1/\omega_3\sin\omega_3\tau \\
-\omega_1\sin\omega_1\tau & 0 & 0 & \cos\omega_1\tau & 0 & 0 \\
0 & -\sqrt{\omega_{\rm in}\omega_{\rm fin}}\sin\phi & 0 & 0 & \sqrt{\frac{\omega_{\rm fin}}{\omega_{\rm in}}}\cos\phi & 0 \\
0 & 0 & -\omega_3\sin\omega_3\tau & 0 & 0 & \cos\omega_3\tau
\end{array}\right)
\end{equation}
where we have set
\begin{equation}
\phi=\frac 23 \tau\frac{\omega_{\rm fin}^2+\omega_{\rm in}\omega_{\rm fin}+\omega_{\rm fin}^2}{\omega_{\rm fin}+\omega_{\rm in}}.
\end{equation}
Equation~\eqref{eq:Uasym} can be interpreted as the evolution operator for the free evolution of oscillators 1 and 3 with their respective frequencies for a time $\tau$ times a squeezing operator dependent on the ratio $\omega_{\rm in}/\omega_{\rm fin}$ combined with a phase space rotation with angle $\phi$.

\section{Definition of Gaussian discord}

In this section we will give a brief mathematical definition of Gaussian discord~\cite{ZurekDiscord2001,HendersonVedral2001, IlluminatiDiscord2004, AdessoDiscord2010, Olivares2012}. For convenience, we use a different ordered basis from the main text, $\tilde{R}=(x_1, p_1, x_2, p_2)$, and recall that the elements of a covariance matrix of two oscillators are defined as $\sigma_{\alpha\beta}=1/2\langle \tilde{R}_\alpha \tilde{R}_\beta +\tilde{R}_\alpha \tilde{R}_\beta\rangle - \langle\tilde{R}_\alpha\rangle \langle\tilde{R}_\beta\rangle$. For a bipartite covariance matrix in this basis,

\begin{equation}
\sigma=\left(\begin{array}{cc}
A&C\\
C^T&B
\end{array}\right),
\end{equation}
where $A$, $B$ and $C$ are $2\times2$ matrices, the matrix $A$ ($B$) relates only to the mean values in subsystem 1 (2) and the matrix $C$ relates to correlations in these mean values. The Gaussian quantum discord of this state can then written as 
\begin{equation}
\mathcal{D}=h\left(\sqrt{I_2}\right)- h(d_-)-h(d_+)+h\left(\sqrt{E_{\rm min}}\right)
\end{equation}
where we have 

\begin{align}
h(x)=&\left(\frac{x+1}{2}\right)\log\left(\frac{x+1}{2}\right)-\left(\frac{x-1}{2}\right)\log\left(\frac{x-1}{2}\right),\\
d_{\pm}^2=&\frac{1}{2}\left(\lambda\pm\sqrt{\lambda^2-4I_4}\right),\\
\lambda=&I_1+I_2+2I_3
\end{align}
for $I_1=\det(A)$, $I_2=\det(B)$, $I_3=\det(C)$ and $I_4=\det(\sigma)$. The quantity $E_{\rm min}$ is given by the following formula:

\begin{equation}
E_{\rm min}=
\begin{cases}
\left(\frac{|I_3|+\sqrt{I_3^2-(I_1-4I_4)(I_2-1/4)}}{2(I_2-1/4)}\right)^2 & \text{if} \hspace{8pt}(I_1 I_2-I_4)^2\leq(I_1+4I_4)(I_2+1/4)I_3^2\\
\\
\frac{1}{I_2}\left(I_1I_2-I_3^2+I_4-\sqrt{(I_1I_2+I_4-I_3^2)^2-4I_1I_2I_4}\right) & \text{otherwise}
\end{cases}.
\end{equation}

\section{Internal energy and quantum correlations for the optimised many-cycle engine}
\begin{figure*}[t!]
\begin{center}
   \includegraphics[width=0.32\textwidth]{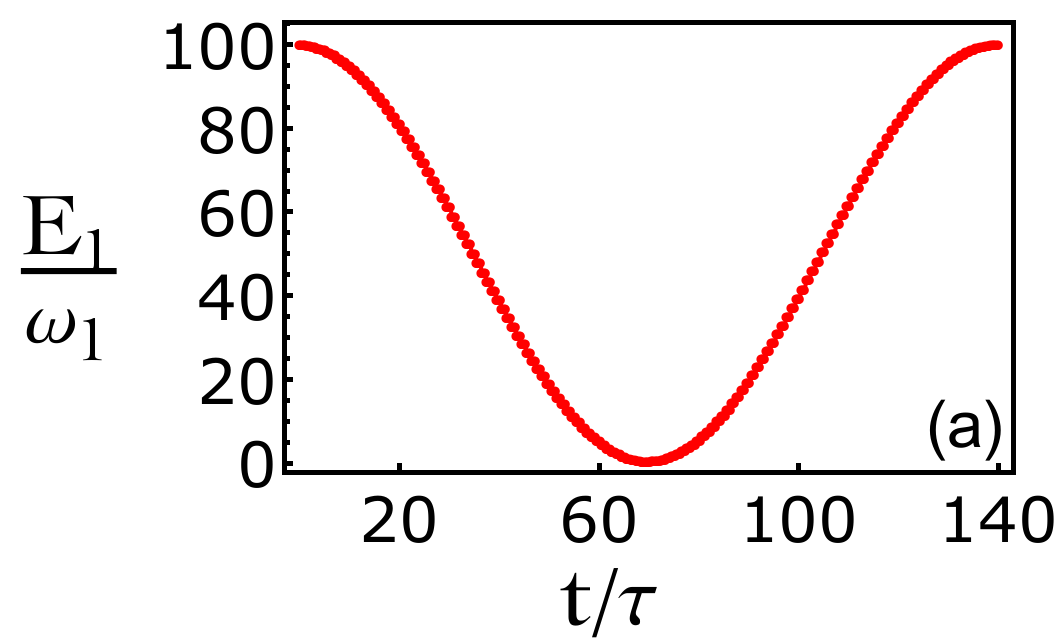}
   \includegraphics[width=0.32\textwidth]{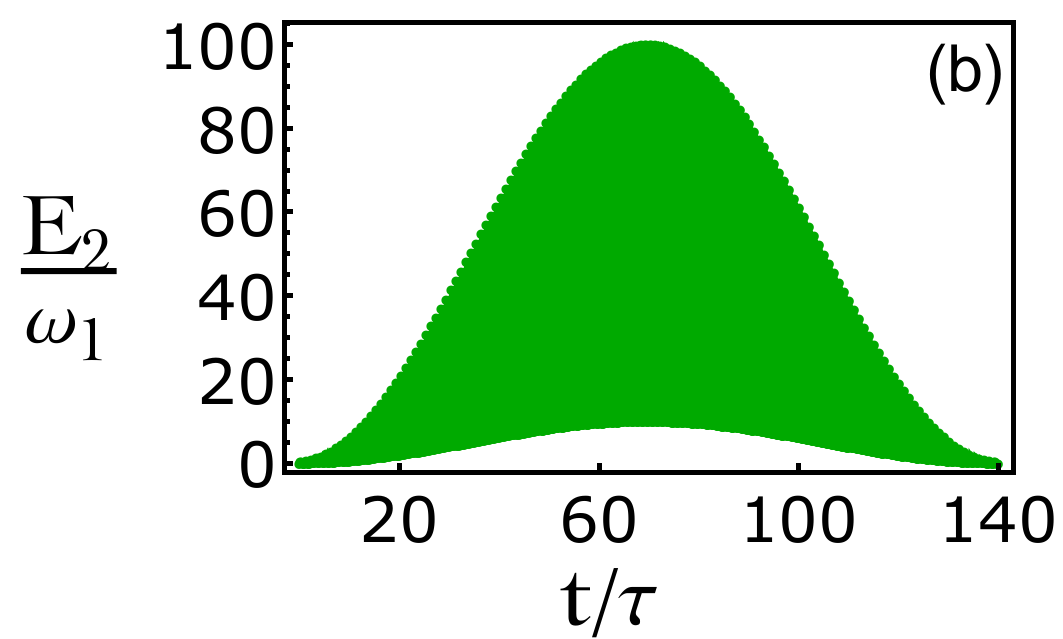}
   \includegraphics[width=0.32\textwidth]{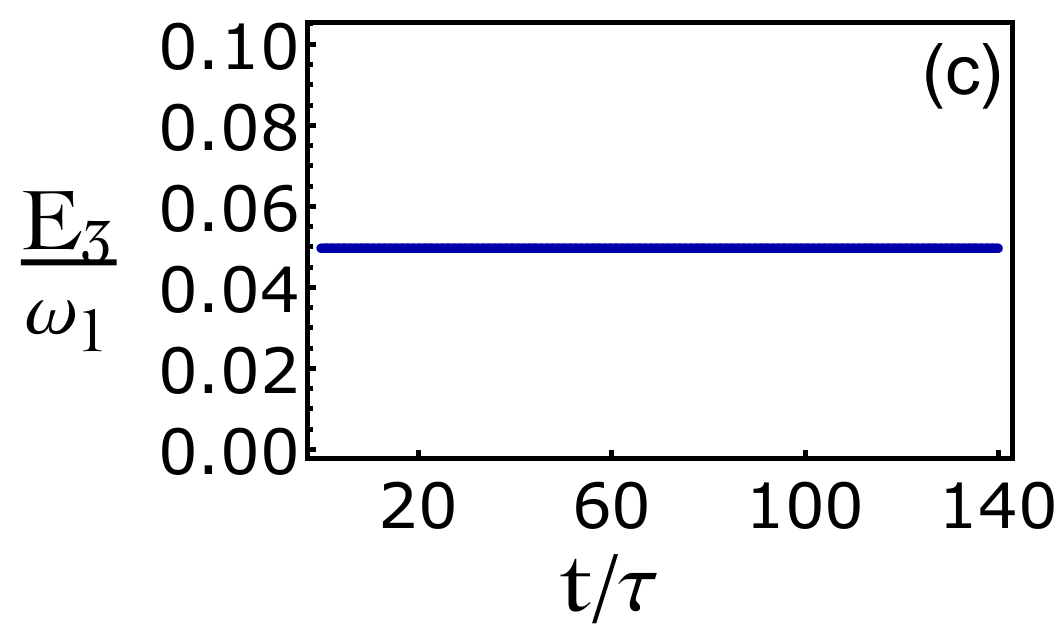}
\caption{  The internal energy dynamics for the first (a), second (b) and third (c) oscillators. }
\label{fig:energyplots}
\end{center}
\end{figure*}

In this section, we present the internal energy dynamics of each oscillator for the optimised many cycle case presented in the main text. We also analyse the discord dynamics between each of the oscillators. The parameters used are $\omega_3=0.1\omega_1, \tau_{\rm comp}=85\omega_1^{-1}, \tau_H=0.59\omega_1^{-1}, \tau_C=0.9996\omega_1^{-1},\alpha_{12}=0.038\omega_1,\alpha_{23}=10^{-4}\omega_1$. Figure \ref{fig:energyplots} shows the internal energy dynamics $E_i=\langle H_i \rangle$ for each oscillator plotted against time in units of the Otto cycle time $\tau=\tau_H+\tau_C+2\tau_{\rm comp}$. The system performs 70 Otto cycles satisfying $\mathcal{W}<0$ and the plots also show the subsequent Otto cycles where the engine reverses. At the end of the 140th Otto cycle each oscillator has almost returned to its initial state. As the coupling strength between the second and third oscillator is extremely weak, the internal energy of the third oscillator does not change. The efficiency of this process is unity, as $Q_2=0$.

\begin{figure}[h!]
    \centering
   \includegraphics[width=0.3\textwidth]{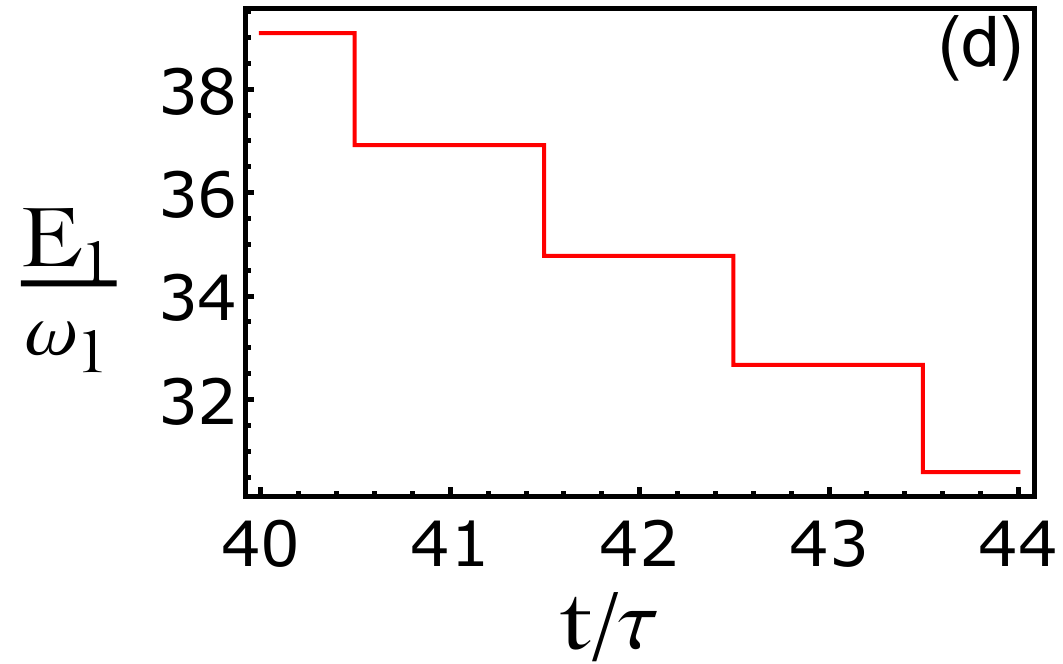}   \includegraphics[width=0.3\textwidth]{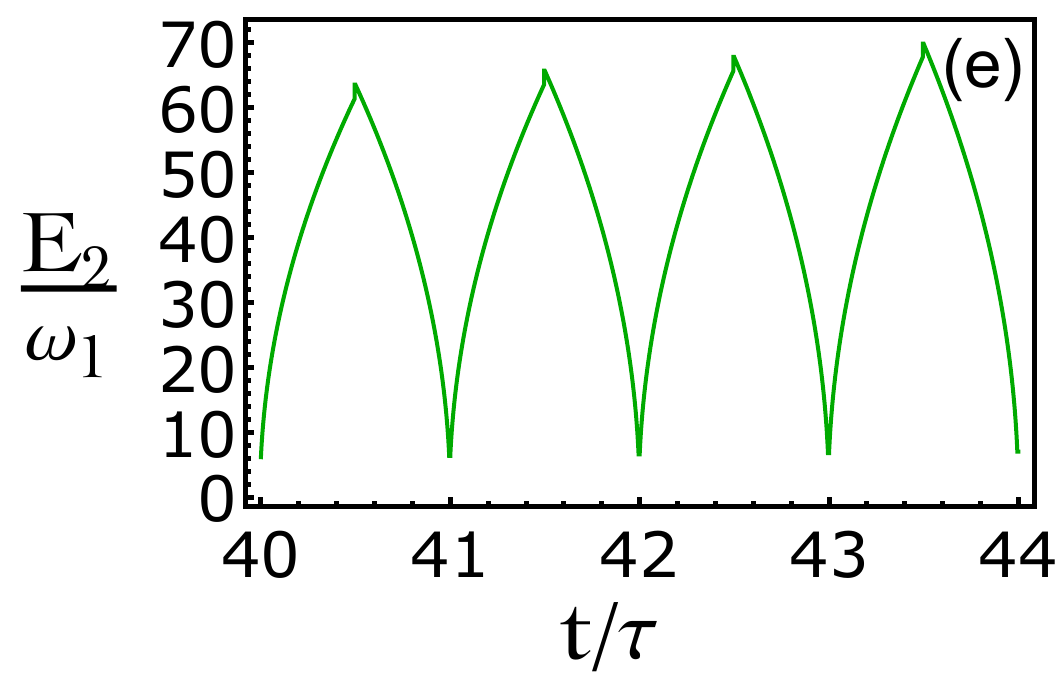}
    \caption{ The internal energies of the first (a) and second (b) oscillators during the 40th to the 44th Otto cycle in Fig. \ref{fig:energyplots}.
 }
    \label{fig:energyplotssub}
\end{figure}

The number of cycles shown in Fig.~\ref{fig:energyplots} makes the fine details of the dynamics difficult to parse. Figure \ref{fig:energyplotssub} shows a closer view of the internal energies of the first and second oscillators from the 40th to the 44th stroke. The internal energy of the first oscillator is shown in Fig.~\ref{fig:energyplotssub}(a). As it is only interacting with the second oscillator during the heating stroke its energy is constant for most of the cycle. Due to the disparity of the time scales of the compression, expansion stages and the heating, cooling stages the latter appear instantaneous on this scale. In Fig.~\ref{fig:energyplotssub}(b) the internal energy of the second oscillator is shown. The large slopes of each arch correspond to the energy increase and decrease caused by the compression and expansion stages. The small peak at the top of each arch corresponds to the heating stroke, where the height of each peak corresponds directly to the height of each energy reduction in panel \ref{fig:energyplotssub}(a).


\end{document}